\documentclass[twocolumn]{aastex7}

\usepackage{amsmath}
\usepackage{CJK}

\newcommand{\bhar}{\ensuremath{\mathrm{\overline{BHAR}}}}

\newcommand{\fagn}{\ensuremath{f_\mathrm{AGN}}}
\newcommand{\lx}{\ensuremath{L_\mathrm{X}}}
\newcommand{\lbol}{\ensuremath{L_\mathrm{bol}}}
\newcommand{\kbol}{\ensuremath{k_\mathrm{X}}}
\newcommand{\plamb}{\ensuremath{p(\lambda|\mstar,z)}}
\newcommand{\pleddrat}{\ensuremath{p(\leddrat|\mstar,z)}}

\newcommand{\lsun}{\ensuremath{L_{\sun}}}
\newcommand{\msun}{\ensuremath{M_{\sun}}}
\newcommand{\mbh}{\ensuremath{M_{\mathrm{BH}}}}
\newcommand{\ledd}{\ensuremath{L_{\mathrm{Edd}}}}
\newcommand{\leddrat}{\ensuremath{\lambda_{\mathrm{Edd}}}}
\newcommand{\leddratlow}{\ensuremath{\lambda_\mathrm{Edd}^\mathrm{low}}}
\newcommand{\leddrathigh}{\ensuremath{\lambda_\mathrm{Edd}^\mathrm{high}}}
\newcommand{\leddratmin}{\ensuremath{\lambda_\mathrm{Edd}^\mathrm{min}}}
\newcommand{\lumcgs}{\ensuremath{\mathrm{erg}\,\mathrm{s}^{-1}}}

\newcommand{\mstar}{\ensuremath{M_\star}}

\newcommand{\nhcgs}{\ensuremath{\mathrm{cm}^{-2}}}

\newcommand{\xray}{\mbox{X-ray}}
\newcommand{\rhobhar}{\ensuremath{\rho_{\mathrm{BHAR}}}}
\newcommand{\z}{\ensuremath{z}}
\newcommand{\phiM}{\ensuremath{\phi_\mathrm{M}}}
\newcommand{\phiMAGN}{\ensuremath{\phi^\mathrm{AGN}_\mathrm{M}}}
\newcommand{\D}{\ensuremath{\mathrm{d}}}
\newcommand{\flambM}{\ensuremath{f(\leddrat,\mstar|z)}}
\newcommand{\nagn}{\ensuremath{n_\mathrm{AGN}}}
\newcommand{\nagneff}{\ensuremath{n_\mathrm{AGN}^\mathrm{eff}}}

\newcommand{\typleddrat}{\ensuremath{\langle\leddrat\rangle}}

\newcommand{\typmbh}{\ensuremath{\langle\mbh\rangle}}
\newcommand{\lbolagn}{\ensuremath{\overline{L_\mathrm{bol}^\mathrm{AGN}}}}
\newcommand{\kev}{\ensuremath{\mathrm{keV}}}
\newcommand{\dex}{\ensuremath{\mathrm{dex}}}
\newcommand{\sfr}{\ensuremath{\mathrm{SFR}}}

\begin{document}
\begin{CJK*}{UTF8}{gbsn}
\title{The Drivers of the Decline in Supermassive Black Hole Growth at $z<2$}

\author[orcid=0000-0002-6990-9058,gname=Zhibo,sname=Yu]{Zhibo Yu (喻知博)}
\affiliation{Department of Astronomy and Astrophysics, The Pennsylvania State University, 525 Davey Lab, University Park, PA 16802, USA}
\affiliation{Institute for Gravitation and the Cosmos, The Pennsylvania State University, University Park, PA 16802, USA}
\email[show]{zvy5225@psu.edu}

\author[orcid=0000-0002-0167-2453]{W. N. Brandt} 
\affiliation{Department of Astronomy and Astrophysics, The Pennsylvania State University, 525 Davey Lab, University Park, PA 16802, USA}
\affiliation{Institute for Gravitation and the Cosmos, The Pennsylvania State University, University Park, PA 16802, USA}
\affiliation{Department of Physics, The Pennsylvania State University, 104 Davey Laboratory, University Park, PA 16802, USA}
\email{wnbrandt@gmail.com}

\author[orcid=0000-0002-4436-6923,gname=Fan,sname=Zou]{Fan Zou}
\affiliation{Department of Astronomy, University of Michigan, 1085 S University, Ann Arbor, MI 48109, USA}
\email{fanzou01@gmail.com}

\author[0000-0002-9036-0063]{Bin Luo}
\affiliation{School of Astronomy and Space Science, Nanjing University, Nanjing 210093, People's Republic of China}
\affiliation{Key Laboratory of Modern Astronomy and Astrophysics (Nanjing University), Ministry of Education, Nanjing 210093, People's Republic of China}
\email{bluo@nju.edu.cn}

\author[0000-0002-8577-2717]{Qingling Ni}
\affiliation{Max-Planck-Institut f\"ur extraterrestrische Physik (MPE), Gie\ss enbachstra\ss e 1, D-85748 Garching bei M\"unchen, Germany}
\email{qingling1001@gmail.com}

\author[0000-0001-7240-7449]{D. P. Schneider}
\affiliation{Department of Astronomy and Astrophysics, The Pennsylvania State University, 525 Davey Lab, University Park, PA 16802, USA}
\affiliation{Institute for Gravitation and the Cosmos, The Pennsylvania State University, University Park, PA 16802, USA}
\email{dps7@psu.edu}

\author[0000-0003-0680-9305]{Fabio Vito}
\affiliation{INAF--Osservatorio di Astrofisica e Scienza dello Spazio di Bologna, Via Gobetti 93/3, I-40129 Bologna, Italy}
\email{fvito.astro@gmail.com}

\begin{abstract}
It is well established that cosmic supermassive black hole (SMBH) growth peaks at $z\approx1.5-2$, followed by a strong decline of $\approx1-1.5\,\dex$ toward the present day, with the comoving number density of higher-luminosity active galactic nuclei (AGNs) peaking at higher redshift (referred to as ``AGN downsizing"). We leverage the best current measurements of the SMBH accretion distribution, based upon data from nine well-characterized extragalactic fields with a ``wedding-cake" design, to investigate and quantify the drivers of the drastic decline in cosmic SMBH growth. The decline in the typical Eddington ratio (\leddrat) of AGNs (decreasing by $\approx1.35\,\dex$ from $z\approx1.5-2$ to $z\approx0.2$) is the dominant driver for the broad decline in SMBH growth, rather than a shift of accretion activity to less-massive SMBHs. As \leddrat\ decreases toward lower redshift, the primary contributor to the cosmic SMBH accretion density (\rhobhar) has shifted from high-\leddrat\ AGNs to low-\leddrat\ AGNs, even though the latter always dominate the comoving AGN number density at $z<4$. We also find that the decline in SMBH growth toward lower SMBH mass in less-massive galaxies is primarily due to the decreasing outburst luminosity rather than the duty cycle.

\end{abstract}


\section{Introduction} \label{sec::intro}

Understanding the growth history of supermassive black holes (SMBHs) is one of the most important topics for extragalactic studies. Observations have revealed tight correlations between black-hole mass (\mbh) and the host-galaxy bulge stellar mass/bulge velocity dispersion \citep[e.g.,][]{Kormendy&Ho+2013}. Furthermore, it has been found that the long-term averaged black-hole accretion rate (\bhar), which is approximated by the sample-averaged BHAR, is correlated with host-galaxy properties such as total stellar mass (\mstar) and bulge star-formation rate (\sfr) over most of cosmic time \citep[e.g.,][]{Xue+2010,Aird+2012,Aird+2018,Yang+2018,Yang+2019,Delvecchio+2020,Zou+2024a}. These relations indicate that SMBHs and host galaxies evolve in a coordinated manner. Therefore, tracing the growth history of SMBHs can provide insights into the mechanism that drives galaxy-SMBH coevolution.

SMBHs in the local universe are generally much more ``quiescent" than their high-redshift counterparts. Observations show that the comoving SMBH accretion-rate density (\rhobhar) traced by the active galactic nucleus (AGN) luminosity function and the comoving number density of AGNs both increase from $z\approx4$ to $z\approx2$ \citep[although the behavior at higher redshift is still not clear, e.g.,][]{Barlow-Hall&Aird+2025}, peak at $z\approx1.5- 2$, and then drop significantly by $\approx1-1.5\,\mathrm{dex}$ by the present day \citep[e.g.,][]{Ueda+2014,Aird+2015,Miyaji+2015,Ananna+2019,Yang+2018,Yang+2023}. Notably, the decline in SMBH growth is also luminosity dependent: the number density of high-luminosity AGNs peaks at a higher redshift ($z\approx2$) than that of lower-luminosity AGNs, which reaches a maximum at $z\approx1$ \citep[e.g.,][]{Cowie+2003,Ueda+2003,Barger+2005,Hasinger+2005,Ueda+2014,Brandt&Alexander+2015,Peca+2023,Alexander+2025}. This phenomenon is referred to as ``AGN downsizing". If AGN luminosity is strictly correlated with \mbh\ (i.e., with fixed Eddington ratio, \leddrat), this result would appear to imply that massive SMBHs largely formed before most less-massive SMBHs, in contrast to the hierarchical formation of dark matter halos based upon the standard cold dark matter model.

This apparent contradiction can be reconciled if we regard AGN downsizing as the dying down of cosmic accretion rather than as a symptom of antihierarchical formation of SMBHs. One proposed explanation is that AGN downsizing may be due to the shift of accretion activity to lower-\mbh\ SMBHs with lower luminosity \citep[e.g.,][]{Heckman+2004}. On the other hand, since it has been shown that \leddrat\ for type~1 quasars has a broad distribution at $z<5$ \citep[e.g.,][]{Babic+2007,Shen&Kelly+2012,Kelly&Shen+2013,Schulze+2015,Suh+2015}, AGN downsizing may also be driven by a decrease in \leddrat\ with low-\leddrat\ AGNs having a greater contribution to \rhobhar\ and \nagn\ at lower redshift. 
Indeed, since the coevolution between SMBHs and galaxies has been established, we should expect the typical SMBH accretion rate to decrease with cosmic time, similar to the observed decline in SFR \citep[e.g.,][]{Madau&Dickinson+2014,Whitaker+2014}. Such coevolution is related to the decreasing gas supply and merger rate toward lower redshift, because galaxy-scale gas inflows and major mergers are known to drive both star-formation and SMBH fueling \citep[e.g.,][]{DiMatteo+2005, Hopkins&Hernquist+2006, Hopkins&Quataert+2010}. It has been suggested that AGN downsizing is partly attributed to the decreasing merger rate toward lower redshift, because merger-induced accretion is thought to produce the most luminous accretion events, causing the earlier peak of \rhobhar\ and \nagn\ for high-luminosity AGNs when major mergers are more frequent \citep[e.g.,][]{Treister+2012}.



The broad decline in SMBH growth at $z\lesssim1.5-2$ is also \mstar-dependent. It has been shown that \bhar, which describes the average accretion rate \textit{per galaxy}, decreases as \mstar\ decreases at fixed redshift \citep[e.g.,][]{Xue+2010,Aird+2012,Aird+2018,Yang+2018,Zou+2024a}. This appears to contrast with the evolution of galaxy SFR at $z\lesssim2$ where star formation preferentially stops in high-mass galaxies \citep[e.g.,][]{Whitaker+2014}, if SMBHs strictly coevolve with their host galaxies. This indicates that SMBH growth is also influenced by physical processes distinct from those driving galaxy-wide star formation. For example, it is likely that more massive galaxies tend to host more massive and luminous SMBHs, and/or the AGN duty cycle (\fagn) typically increases with \mstar\ thereby increasing \bhar\ \citep[e.g.,][]{Aird+2012,Bongiorno+2012,Heckman&Best+2014,Zou+2024a}.

\xray\ surveys arguably provide the most robust constraints on SMBH growth because \xray\ emission has reduced bias due to its high penetrating power and the large contrast between the \xray\ emission of AGNs and stellar components \citep[e.g.,][]{Brandt&Yang+2022}. Recently, \citet{Zou+2024a} have obtained the best measurements of \bhar\ (traced by X-ray emission) utilizing the superb \xray\ and multiwavelength data in nine well-studied extragalactic surveys. In this work, we quantitatively investigate the decline in SMBH growth at $z\lesssim2$ based upon the results of \citet{Zou+2024a}, and address the following key questions:
\begin{enumerate}
    \item At $z\lesssim2$, as redshift decreases, what causes the broad decline in SMBH growth? Is it due to (a) a shift of accretion activity to lower \mbh\ in generally lower-\mstar\ galaxies, (b) reduction in the typical \leddrat\ at the same \mbh, (c) reduction in the number density of AGNs, or (d) some combination of these possibilities?
    \item Why does SMBH growth decline as \mstar\ decreases? Does \mstar\ mainly modulate the typical outburst luminosity or duty cycle to reduce SMBH growth?
\end{enumerate}


This paper is structured as follows. Section~\ref{sec::methodology} describes the data and methodology. Sections~\ref{subsec::decline_in_z} and \ref{subsec::mstar} present our results for the two key questions, respectively. Section~\ref{sec::summary} summarizes this work. Throughout the paper,  we adopt a flat $\Lambda$CDM cosmology with $H_0=70\,\rm km\,s^{-1}\,Mpc^{-1}$, $\Omega_\Lambda=0.70$, and $\Omega_\mathrm{M}=0.30$.

We denote the $\mbh-\mstar$ relation as $\mbh=\eta(\mstar)$. Most of our analyses adopt a linear relation with $\eta(\mstar)=0.002\,\mstar$ \citep{Marconi+2004} unless otherwise specified. This linear relation is similar to the one in \citet{Reines&Volonteri+2015} that is based upon dynamically measured \mbh. It has been shown that the $\mbh-\mstar$ relation does not have a significant redshift evolution at $z<2$ \citep[e.g.,][]{Suh+2020,Li+2023}. All $\mbh-\mstar$ relations typically have $\approx0.5\,\dex$ scatters, and our relation is appropriate in an average sense.

\section{Data and Methodology} \label{sec::methodology}

\subsection{Data} \label{subsec::data}

The best current measurements of \bhar\ by \citet{Zou+2024a} utilize data from nine well-studied extragalactic surveys. These surveys follow a standard ``wedding-cake" design and consist of deep, pencil-beam and shallower, wider surveys (spanning $0.05-60\,\mathrm{deg^2}$), allowing us to effectively explore a wide range of parameter space. The fields include four of the Cosmic Assembly Near-infrared Deep Extragalactic Legacy Survey (CANDELS) fields, four of the Vera C. Rubin Observatory Legacy Survey of Space and Time (LSST) Deep-Drilling Fields (DDFs), and the eROSITA Final Equatorial Depth Survey (eFEDS) field. These fields have sensitive \xray\ coverage and superb multiwavelength data, providing quality characterization of $\approx8000$ \xray-selected AGNs and 1.3 million galaxies that are above the \mstar-completeness limits. We summarize the information for these fields as follows:
\begin{enumerate}
\item CANDELS: We use the ultra-deep \xray-to-infrared data in four of the CANDELS \citep{Grogin+2011,Koekemoer+2011} fields: GOODS-S, GOODS-N, Extended Groth Strip (EGS), and UKIRT Infrared Deep Sky Survey Ultra-Deep Survey (UDS). These fields have ultra-deep Chandra observations reaching megasecond depths that allow us to effectively sample AGNs at high redshift and/or low luminosity: \citet{Luo+2017} for GOODS-S, \citet{Xue+2016} for GOODS-N, \citet{Nandra+2015} for EGS, and \citet{Kocevski+2018} for UDS. The galaxy catalog is from \citet{Yang+2019}, in which the host-galaxy properties of \mstar\ and SFR are derived using SED-fitting.
\item LSST DDFs: We use four of the LSST DDFs \citep{Brandt+2018,Zou+2022}: Cosmic Evolution Survey (COSMOS), Wide Chandra Deep Field-South (W-CDF-S), European Large-Area Infrared Space Observatory Survey-S1 (ELAIS-S1), and XMM-Newton Large Scale Structure (XMM-LSS). These fields have sensitive multiwavelength data, including medium-depth X-ray coverage. The fifth LSST DDF, Euclid Deep Field-South (EDF-S), was recently selected as one of the DDFs in 2022, and it currently lacks data of comparable quality, so we did not include it. For COSMOS, Chandra provides $\approx160\,\mathrm{ks}$ \xray\ depth \citep{Civano+2016}, and the galaxy properties are provided in the Appendix of \citet{Yu+2023} derived from SED-fitting. For the other three DDFs, the \xray\ coverage is provided by the XMM-Spitzer Extragalactic Representative Volume Survey (XMM-SERVS). \citet{Ni+2021} provide $\approx30\,\mathrm{ks}$ XMM-Newton coverage for W-CDF-S and ELAIS-S1. \citet{Chen+2018} provide $\approx40\,\mathrm{ks}$ XMM-Newton coverage for XMM-LSS. The galaxy properties of these three fields are provided by \citet{Zou+2022}. Note that we only focus on the regions that have both \xray\ and sensitive near-infrared coverage, totaling $\approx13\deg^2$. We have ensured that regions with overlap with CANDELS are not double-counted.
\item eFEDS: We focus on the $60\deg^2$ GAMA09 subfield inside the full eFEDS field because its good multiwavelength coverage allows us to constrain host-galaxy properties. eFEDS has been observed by eROSITA with $\approx2\,\mathrm{ks}$ depth, most sensitively at $<2.3\,\kev$. Due to the soft \xray\ coverage, the \xray\ properties are from \citet{Liu+2022}, who performed \xray\ spectral analyses to account for obscuration effects. The host-galaxy properties are provided in \citet{Yu+2023}.
\end{enumerate}

For further information on these fields, see Section~2 of \citet{Zou+2024a}. 
The wedding-cake design of the fields and their quality source characterization enabled us to sample AGNs with a wide range of rest-frame $2-10\,\kev$ intrinsic luminosity (\lx; $\log[\lx/\lumcgs]\approx40-45$) across $z=0-4$, so that we can capture most of the SMBH growth at $z<4$. Particularly relevant to this work are the significantly tighter constraints set by \citet{Zou+2024a} upon SMBH growth at $z\lesssim1$ for a wide range of \mstar, which allow us to quantify the decline in SMBH growth accurately. This aspect is mostly enabled by the large sampled volume of the $13\deg^2$ XMM-SERVS and $60\deg^2$ eFEDS surveys.

To measure SMBH growth from \xray\ surveys, it is necessary to correct for the obscured accretion power. This is particularly important for eFEDS, since its soft \xray\ coverage can lead to a higher fraction of missed obscured AGNs, thereby introducing bias into the results. The correction utilizes the detection probability of our \xray\ surveys to correct for the obscured accretion power, which is calibrated based upon the well-determined $\log N-\log S$ relation, the expected surface number density per unit \xray\ flux with the detection procedures deconvolved \citep[see Section~3.1.1 of ][]{Zou+2024a}. Thus, the measurements have been corrected for all of the missed obscured Compton-thin (CN) AGNs and for part of the missed Compton-thick (CT) AGNs in our fields. \citet{Zou+2024a} showed that excluding eFEDS from the analyses leads to differences in \bhar\ smaller than the $1\sigma$ statistical uncertainties. We have further verified in Appendix~\ref{appedix:noefeds} that the median values of our results remain consistent when eFEDS is excluded. Importantly, we do not rely solely on eFEDS to probe SMBH growth at low redshift and/or high luminosity, as the $\approx13\,\deg^2$ area in the LSST DDFs with sensitive \xray\ coverage above $2\,\kev$ already provides meaningful constraints. eFEDS increases the number of our \xray\ AGNs at $z\lesssim1$ by $\approx60\%$ and thereby reduces statistical uncertainties, allowing more accurate measurements of the SMBH growth decline since $z\approx2$.

One systematic uncertainty arises from the fact that correcting for CT accretion is generally difficult for our fields, particularly for eFEDS due to its soft \xray\ coverage. However, the missing CT accretion should not have a material impact on our results for two reasons. First, some CT AGNs can still be detected by Chandra or XMM-Newton \citep[e.g.,][]{Li+2020,Yan+2023}, particularly since increasing redshift allows us to probe up to rest-frame $10-25\,\kev$ at $z<2$ with greater penetrating power. Second, using the \xray\ luminosity function (XLF) with the column-density distribution from Sections~3 and 6 of \citet{Ueda+2014}, \citet{Zou+2024a} found that the fractional accretion power at column densities above $10^{24}\,\nhcgs$ is $\approx38\%$ across all redshifts; similarly, \citet{Buchner+2015} showed that the \xray\ luminosity density at $\log[\lx/\lumcgs]>43.2$ contributed by CT AGNs is $\approx35-45\%$ at $z=0.8-3.6$. Thus, the systematic bias upon SMBH growth from potentially missed CT accretion is $\lesssim0.2\,\dex$, with the $0.2\,\dex$ upper limit corresponding to the extreme case in which all of the $\approx40\%$ CT accretion power is missed.\footnote{The $\approx0.2\,\dex$ systematic bias has been estimated by assuming that we could only detect the $\approx60\%$ contribution from CN accretion: $\log(0.6)=-0.22\,\dex$.} This $\lesssim0.2\,\dex$ bias arises solely from possible missed CT accretion after applying the obscuration correction of \citet{Zou+2024a}, and is not specifically attributed to eFEDS.


\subsection{Methodology}
 
We denote \plamb\ and \pleddrat\ as the conditional probability density per unit $\log\lambda$ and $\log\leddrat$ of a galaxy with (\mstar, $z$) hosting an AGN with $\lambda\equiv\lx/\mstar/[\lumcgs\msun^{-1}]$ and $\leddrat\equiv\lbol/\ledd$, respectively, where \lbol\ is the AGN bolometric luminosity and \ledd\ is the Eddington luminosity of the SMBH. \citet{Zou+2024a} obtained \plamb\ at $z=0-4$ with $\log\mstar=9.5-12$, reaching down to $\log\lambda>31.5$. The \xray\ bolometric correction factor as a function of \lbol\ (in units of \lumcgs) is $\kbol(\lbol)$. For a fixed \mstar, by definition, 
\begin{equation}
\begin{split}
\pleddrat&=\plamb\frac{\D\log\lambda}{\D\log\leddrat}\\
&=\plamb\left(1-\frac{\D\log\kbol(\lbol)}{\D\log\lbol}\right).
\end{split}
\end{equation}
The second equality holds because
\begin{equation}
\begin{split}\label{eqn::derivation}
\leddrat&=\frac{\lx\kbol(\lbol)}{1.26\times10^{38}\times\mbh}=\frac{\lx\,\kbol(\lbol)\,\mstar}{1.26\times10^{38}\times\eta(\mstar)\,\mstar}\\
&=\lambda\,\kbol(\lbol)\,\frac{\mstar}{1.26\times10^{38}\times\eta(\mstar)}.
\end{split}
\end{equation}
The \mstar-related term is constant when we calculate \pleddrat. We take the logarithm of both sides of Equation~\ref{eqn::derivation}, and differentiate both sides with respect to $\log\leddrat$. Since $\D\log\leddrat=\D\log\lbol$ when $\eta(\mstar)$ is fixed, we have
\begin{equation}
    \frac{\D\log\lambda}{\D\log\leddrat}=1-\frac{\D\log\kbol(\lbol)}{\D\log\lbol}.
\end{equation}
We adopt the $\log\kbol-\log\lbol$ relation from Equation~2 of \citet{Duras+2020}:
\begin{equation}
    \kbol=10.96\left[1+\left(\frac{\log[\lbol/\lsun]}{11.93}\right)^{17.79}\right].
\end{equation}
\kbol\ diverges at very high \lbol, so we set an upper limit of $\kbol=100$ because this is about the maximum value observed in previous literature \citep[e.g.,][]{Marconi+2004,Hopkins+2007,Vasudevan+2007}. The limit corresponds to large \lbol\ and \lx\ thresholds of $\log[\lbol/\lumcgs]=47.0$ and $\log[\lx/\lumcgs]=45.0$. For \lbol\ higher than the threshold \lbol, we apply a constant $\kbol=100$. Although higher \kbol\ has been observed for very luminous quasars \citep[e.g.,][]{Martocchia+2017}, these quasars are very rare and are generally missed in our fields. We have tested that setting the upper limit of \kbol\ to $\approx400$ only changes our results for the highest \leddrat\ bin (Section~\ref{subsec::decline_in_z}) by $<0.2\,\dex$.

With \pleddrat\ [or equivalently, \plamb], we can derive many metrics describing SMBH growth. For example, one of the key results in \citet{Zou+2024a} is the measurement of \bhar, sampled down to AGNs with $\lambda_\mathrm{Edd}=0.0013$ (corresponding to $\log\lambda=31.5$). We adopt a radiative efficiency of $\epsilon=0.1$, which is a typical value for the general AGN population. Although at $\leddrat\lesssim0.01$, radiatively inefficient accretion flows \mbox{(RIAFs)} may become relevant, they are not expected to contribute a significant fraction of SMBH growth \citep[e.g.,][]{Narayan&Yi+1995,Yuan&Narayan+2014}. In fact, we have checked that if we only focus on AGNs with $\leddrat>0.01$, our results are not materially different. For SMBHs accreting between \leddratlow\ and \leddrathigh, \bhar\ can be expressed as
\begin{equation}
\begin{split}
&\bhar(\mstar, z; \leddratlow, \leddrathigh)\\
&=\int_{\log\leddratlow}^{\log\leddrathigh}\frac{(1-\epsilon)\leddrat\mstar\,\kappa(\mstar)}{\epsilon\,c^2}p(\leddrat|\mstar,z)\D\log\leddrat,
\end{split}
\end{equation}
where 
\begin{equation}\label{eqn::kappa}
    \kappa(\mstar)=[1.26\times10^{38}\times\eta(\mstar)/\mstar]\,\lumcgs\msun^{-1},
\end{equation}
which converts $\leddrat\mstar$ to \lbol. For most of our analyses, under the assumption of $\eta(\mstar)\equiv\mbh=0.002\,\mstar$, $\kappa(\mstar)$ is constant. However, we retain the notation in Equation~\ref{eqn::kappa} because it is applicable for other nonlinear $\mbh-\mstar$ relations \citep[e.g.,][]{Reines&Volonteri+2015,Greene+2020}. It is worth noting that \bhar\ represents the actual \textit{growth rate} of SMBHs rather than the mass accretion rate at a certain radius from the SMBH. In the following Section, we will derive other quantities of interest using \pleddrat, including \rhobhar\ and \nagn, to answer our key questions in Section~\ref{sec::intro}. We derive the statistical uncertainties of our results using a Monte Carlo method. \citet{Zou+2024a} obtained 10,000 posterior samples of \plamb\ with a Hamiltonian Monte Carlo sampler, adopting parameter priors described in Section~3.1.3 of their work. From these posterior samples, we randomly select 1000 samples without replacement and repeat our analyses (Section~\ref{sec::results}) 1000 times to estimate the statistical uncertainties. We do not consider additional sources of uncertainty beyond \plamb\ (e.g., \kbol, $\mbh-\mstar$ relation). While this approach does not affect the median values of our results, it may introduce an additional uncertainty of $\approx0.2-0.3\,\dex$. Unless otherwise specified, all quoted statistical uncertainties correspond to 90\% confidence intervals.


\section{Results} \label{sec::results}

\subsection{What Primarily Causes the Decline in SMBH Growth as a Function of Redshift?} \label{subsec::decline_in_z}

In this subsection, we investigate the reason why \rhobhar\ declines dramatically at $z\lesssim1.5-2$. As a useful first-order approximation, \rhobhar\ can be factored into three redshift-dependent components and a constant factor:
\begin{equation}\label{eqn::rho_decompose}
\rhobhar=\nagneff\times\langle\leddrat\rangle\times\langle\mbh\rangle\times \frac{(1-\epsilon)\times1.26\times10^{38}}{\epsilon c^2},
\end{equation}
where \typleddrat\ and \typmbh\ represent the typical \leddrat\ and typical \mbh\ of AGNs that contribute to most of \rhobhar. Note that $1.26\times10^{38}\typmbh=\langle\mstar\kappa(\mstar)\rangle$. The last three terms on the right-hand side of Equation~\ref{eqn::rho_decompose}, $\typleddrat\typmbh(1-\epsilon)/\epsilon c^2\times1.26\times10^{38}$, simply describe the typical SMBH accretion rate \textit{per AGN}. \nagneff\ is the ``effective" AGN number density set as if all AGNs have the same SMBH mass of \typmbh\ and accrete at \typleddrat\ in order to produce the observed \rhobhar. \nagneff, \typleddrat, and \typmbh\ correspond to the three redshift-dependent factors that control the decline in SMBH growth in key question~1 in Section~\ref{sec::intro}. While \typleddrat\ and \typmbh\ represent the properties of a typical AGN, \nagneff\ is useful in assessing the expected number density of those AGNs. By definition, \nagneff\ should only be calculated using Equation~\ref{eqn::rho_decompose} after \rhobhar, \typleddrat, and \typmbh\ are determined. Note that \nagneff\ may be different from the true AGN number density \nagn, because \typleddrat\ and \typmbh\ are not the arithmetic average of \leddrat\ and \mstar\ for all AGNs. We will further illustrate this point in Section~\ref{subsubsec::impact_of_three}. In the following subsections, we aim to measure \rhobhar, \typleddrat, \typmbh, and \nagneff\ in different redshift bins. The redshift bins are based upon those in \citet{Weaver+2023}, with bin boundaries at 0.2, 0.5, 0.8, 1.1, 1.5, 2.0, 2.5, 3.0, 3.5, and 4.0 (i.e., this is the binning available for the galaxy stellar mass function; see Section~\ref{subsubsec::rhobhar}). The numbers of \xray\ AGNs and normal galaxies are summarized at the bottom of Table~\ref{tab::decline}. We also aim to quantify the impact of these factors on the decline of \rhobhar, i.e.,
\begin{equation}
\Delta\log\rhobhar=\Delta\log\nagneff+\Delta\log\typleddrat+\Delta\log\typmbh,
\end{equation}
where the $\Delta$'s represent the change of these quantities from the $z=1.5-2$ bin to $z=0.2-0.5$ bin.

\subsubsection{Cosmic BHAR Density and AGN Number Density}\label{subsubsec::rhobhar}

We can calculate \rhobhar\ in a redshift bin for SMBHs accreting between \leddratlow\ and \leddrathigh\ by convolving \bhar\ with the galaxy stellar mass function (SMF; \phiM) in that redshift bin:

\begin{equation}\label{eqn::rhobhar}
\begin{split}
&\rhobhar(z;\leddratlow,\leddrathigh)\\
&=\int_{9.5}^{12}\bhar(\mstar,\z;\leddratlow,\leddrathigh)\phiM(\mstar|z)\D\log\mstar\\
&=\int_{9.5}^{12}\int_{\log\leddratlow}^{\log\leddrathigh}\frac{(1-\epsilon)\pleddrat\phiM(\mstar|z)}{\epsilon\,c^2}\\
&\phantom{A The spaceholder}\times\leddrat\mstar\kappa(\mstar)\,\D\log\leddrat\D\log\mstar\\
&=\frac{1-\epsilon}{\epsilon\,c^2}\int_{9.5}^{12}\int_{\log\leddratlow}^{\log\leddrathigh}f(\leddrat,\mstar|z)\\
&\phantom{A The spaceholder}\times\leddrat\mstar\kappa(\mstar)\,\D\log\leddrat\D\log\mstar,
\end{split}
\end{equation}
where 
\begin{equation} \label{eqn::flambM}
\flambM\equiv\pleddrat\phiM(\mstar|z), 
\end{equation}
We adopt the SMF from \citet{Weaver+2023}, using the same redshift bins as defined in their work. Our analyses only consider SMBHs residing in galaxies with $\log\mstar=9.5-12$ because the \bhar\ for massive black holes in dwarf galaxies is still poorly understood, and dwarf galaxies do not necessarily always host massive black holes \citep[e.g.,][]{Miller+2015, Gallo&Sesana+2019,Zou+2025}.

\begin{figure}[t]\centering
\includegraphics[width=\linewidth]{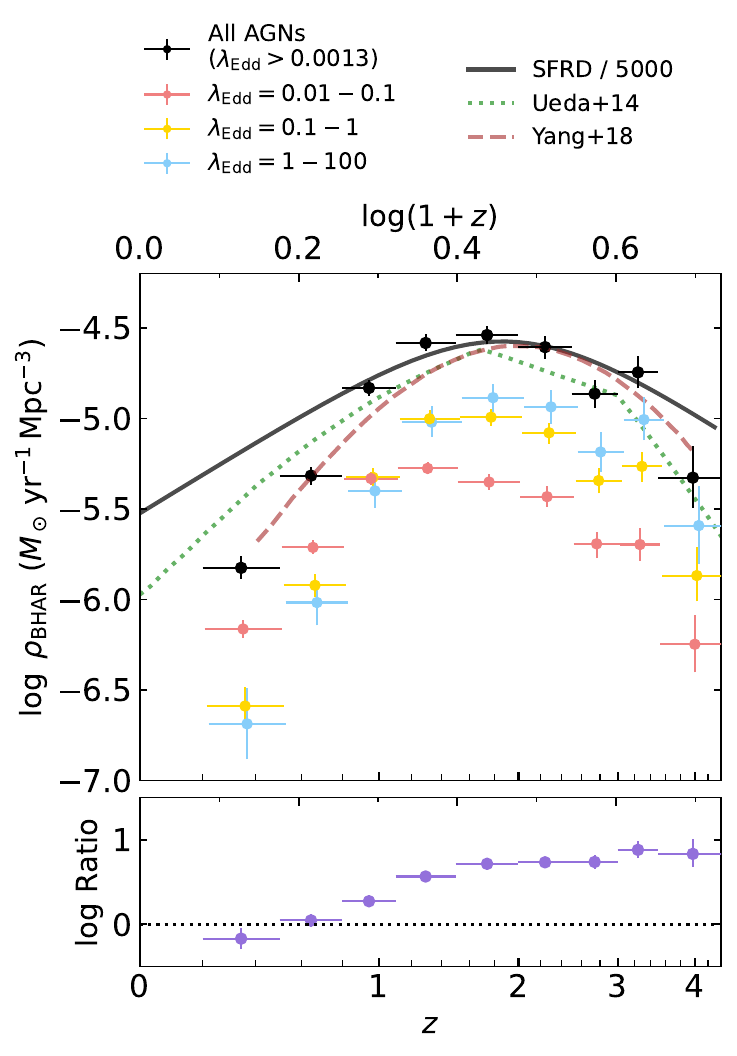}
\caption{Top panel: \rhobhar\ as a function of redshift. The black data points represent the total \rhobhar\ sampled by our data. The red, yellow, and blue data points represent \rhobhar\ for AGNs accreting at $\leddrat=0.01-0.1$, $\leddrat=0.1-1$, and $\leddrat=1-100$, respectively. For comparison, the cosmic SFRD \citep{Madau&Dickinson+2014} scaled by a factor of 5000 is shown as the dashed line to approximate the peak of \rhobhar\ at $z\approx2$. The green dotted line and the brown dashed line represent the total \rhobhar\ from \citet{Ueda+2014} and \citet{Yang+2018}, respectively. Bottom panel: logarithm of the ratio of \rhobhar\ contributed by AGNs with $\leddrat>0.1$ to that contributed by AGNs with $\leddrat<0.1$. The error bars represent the 90\% confidence intervals derived with our Monte Carlo method.}\label{fig::rhobhar}
\end{figure}

The top panel of Figure~\ref{fig::rhobhar} displays \rhobhar\ for AGNs in different \leddrat\ bins as functions of redshift. For the total \rhobhar\ sampled by all of our AGNs, we successfully produce the peak at $z\approx1.5-2$, and the decline of \rhobhar\ from $z=1.5-2$ to $z=0.2-0.5$ is $-1.28_{-0.08}^{+0.08}\,\mathrm{dex}$. In Appendix~\ref{appedix:functional_fits}, we present simple functional fits to our measurements in different \leddrat\ bins. For comparison, in Figure~\ref{fig::rhobhar} we present the cosmic star-formation rate density (SFRD) from \citet{Madau&Dickinson+2014}, which is scaled by a factor of 5000 to approximately match the peak of total \rhobhar. The decline in \rhobhar\ is larger than the decline in SFRD, which is consistent with previous results \citep[e.g.,][]{Aird+2015, Yang+2023}. We also show the total \rhobhar\ from \citet{Ueda+2014} and \citet{Yang+2018}. Our results are generally in agreement with theirs except at $z<0.8$ where our total \rhobhar\ is slightly lower than that in \citet{Ueda+2014}. This may be because \citet{Ueda+2014} adopted the \xray\ bolometric correction in \citet{Hopkins+2007}, which is slightly larger than that in \citet{Duras+2020} at $\log\lx=42-45$.

Our results also show AGN downsizing in different \leddrat\ bins instead of the typical downsizing phenomena observed in different luminosity bins. We find that \rhobhar\ contributed by high-\leddrat\ AGNs peaks at a higher redshift than for low-\leddrat\ AGNs. The bottom panel of Figure~\ref{fig::rhobhar} shows the ratio of \rhobhar\ contributed by AGNs with $\leddrat>0.1$ to that by AGNs with $\leddrat<0.1$. At $z\gtrsim1$, \rhobhar\ is mainly contributed by high-\leddrat\ AGNs. Our results also indicate that super-Eddington accretion contributes to much of the accretion density at $z\gtrsim2$, but the difference is not very significant and may suffer additional bias. This is because the difference between the $\leddrat=0.1-1$ bin and $\leddrat=1-100$ bin is generally comparable to the 90\% statistical uncertainty, and at $z>3$, there may be more hidden accretion power at low-$\leddrat$ than expected (see the discussion at the end of Section~\ref{subsubsec::impact_of_three}). At $z\lesssim0.5$, more of the accretion activity has shifted to low-\leddrat\ AGNs, and about half of the \rhobhar\ at $z=0.2-0.5$ is contributed by AGNs with $\leddrat<0.1$. We have also verified that our result for super-Eddington AGNs is not sensitive to the choice of \leddrathigh. The results are very similar when $\leddrathigh$ takes any value $>100$; the difference is $\lesssim0.15\,\mathrm{dex}$ when we change $\leddrathigh$ from 100 to 10. This difference is also smaller than the 90\% statistical uncertainty for the $\leddrat=1-100$ bin. So far, we have not yet quantified the decline in \typleddrat, but the fact that the primary contributor to \rhobhar\ shifts from AGNs with $\leddrat>0.1$ to those with $\leddrat<0.1$ indicates that \typleddrat\ has declined significantly from $z\approx2$ to $z\approx0.2$. We have verified that the results for \rhobhar\ remain similar after excluding eFEDS from our analyses in Appendix~\ref{appedix:noefeds}.

\begin{figure*}[htbp]
\centering
\includegraphics[width=\linewidth]{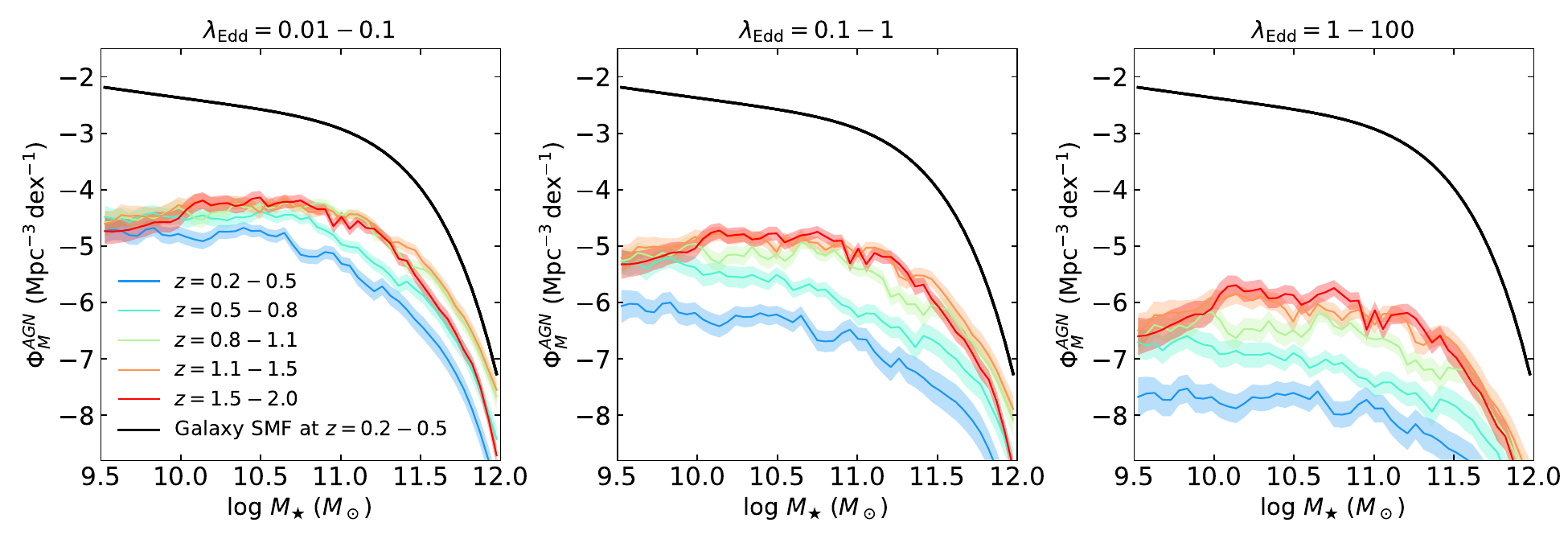}
\caption{AGN host-galaxy SMF \phiMAGN\ at different redshifts, with colors defined in the legend. The left, middle, and right panels show the $\leddrat=0.01-0.1$, $\leddrat=0.1-1$, and $\leddrat=1-100$ bins, respectively. For comparison, the galaxy SMF  at $z=0.2-0.5$ in \citet{Weaver+2023} is shown as black solid curves. The colored shaded stripes represent the $1\sigma$ uncertainty from \pleddrat.}\label{fig::phiMAGN}
\end{figure*}

Similar to the calculation of \rhobhar, the AGN host-galaxy SMF (\phiMAGN) for AGNs accreting between \leddratlow\ and \leddrathigh\ can be directly calculated by convolving \pleddrat\ with \phiM:
\begin{equation} \label{eqn::phiAGN}
\begin{split}
&\phiMAGN(\mstar;\leddratlow,\leddrathigh|z)\\
&=\int_{\log\leddratlow}^{\log\leddrathigh}\pleddrat\phiM(\mstar|z)\D\log\leddrat\\
&=\int_{\log\leddratlow}^{\log\leddrathigh}\flambM\D\log\leddrat.
\end{split}
\end{equation}
The \phiMAGN\ at $z<2.0$ in different \leddrat\ bins is shown in Figure~\ref{fig::phiMAGN}. From $z=1.5-2.0$ to $z=0.2-0.5$, the decline in \phiMAGN\ is most significant in the $\leddrat=1-100$ bin, reaching $\approx2\,\dex$ at $\log\mstar=10-11.5$, while the decline in the $\leddrat=0.01-0.1$ bin is only $<1\,\dex$ at all \mstar. The result is consistent with that in Figure~\ref{fig::rhobhar}, where \rhobhar\ for high-\leddrat\ AGNs declines more significantly than that for low-\leddrat\ AGNs.

With \phiMAGN\, we can then further derive \nagn\ for AGNs accreting between \leddratlow\ and \leddrathigh\ as
\begin{equation} \label{eqn::nagn}
\begin{split}
    &\nagn(z;\leddratlow, \leddrathigh)\\
    &=\int_{9.5}^{12}\phiMAGN(\mstar;\leddratlow,\leddrathigh|z)\D\log\mstar.
\end{split}
\end{equation}
Note that Equation~\ref{eqn::nagn} is not intended to calculate \nagneff, because by definition, after quantifying \typleddrat\ and \typmbh, \nagneff\ is obtained using Equation~\ref{eqn::rho_decompose}. However, Equation~\ref{eqn::nagn} can still provide insights into how many AGNs have \leddrat\ and \mbh\ similar to \typleddrat\ and \typmbh. In the next subsection, we will first quantify \typleddrat\ and \typmbh. With them, we can further determine \nagneff.

\subsubsection{Quantifying the Impact of \typleddrat, \typmbh, and \nagneff}\label{subsubsec::impact_of_three}

The quantities \typleddrat\ and \typmbh\ are less well-defined than \rhobhar\ and \nagn. Since we expect \typleddrat\ and \typmbh\ can represent the \leddrat\ and \mbh\ of the AGNs that contribute to most of \rhobhar, we can determine \typleddrat\ and \typmbh\ by examining the contribution to \rhobhar\ in different ($\log\leddrat$, $\log\mbh$) grid cells in the $\log\leddrat-\log\mbh$ plane and identifying the region that produces most of the contribution.

\begin{figure*}[ht]
\centering
\includegraphics[width=\linewidth]{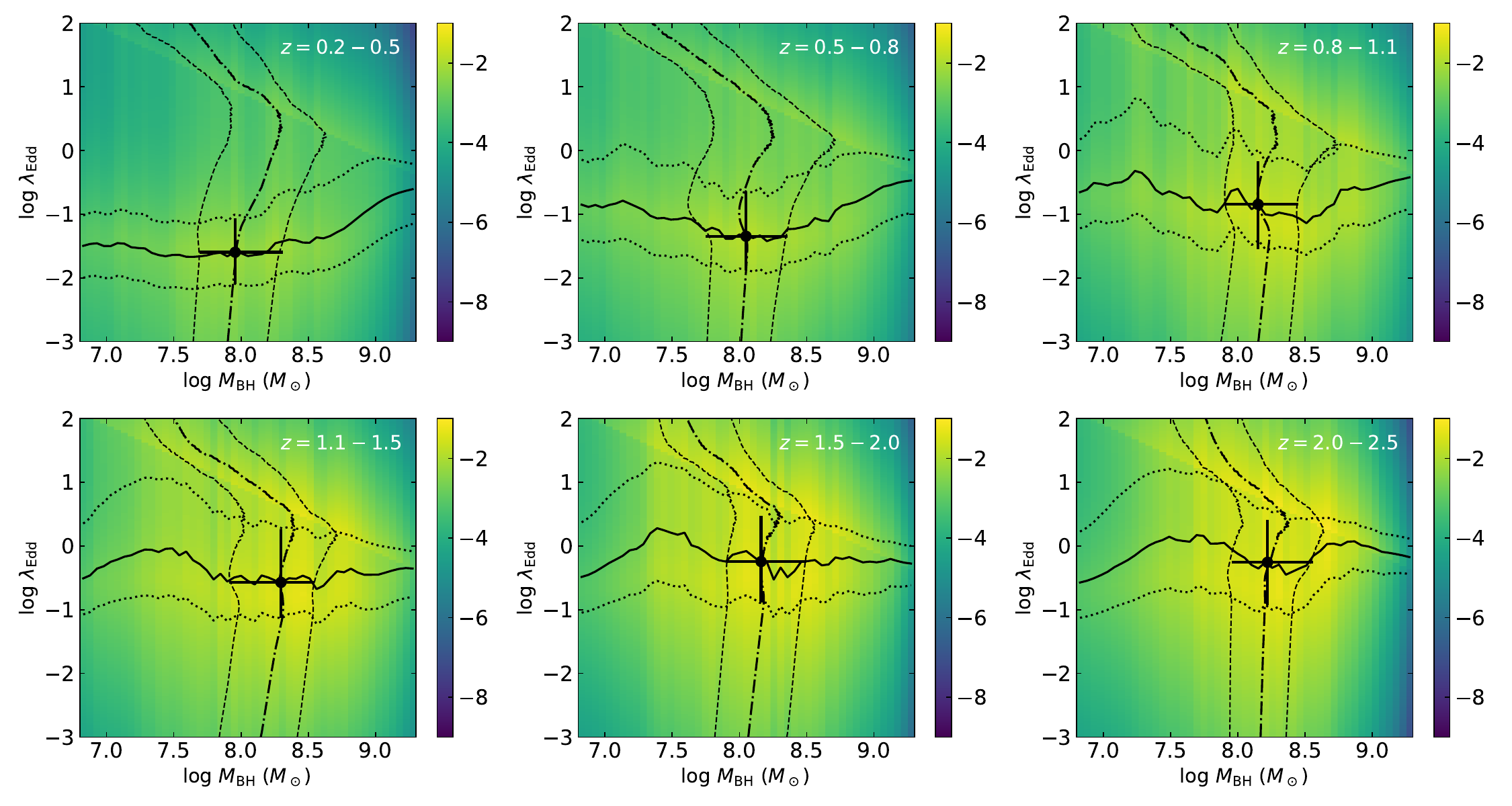}
\caption{Contribution to \rhobhar\ per galaxy at different redshifts assuming the median \pleddrat\ and $\mbh=0.002\mstar$. The color bars represent the logarithm of $(1-\epsilon)/(\epsilon c^2)\times F(\leddrat,\mstar|z)\times\mstar\leddrat\kappa(\mstar)$. The median values of $\log\leddrat$ ($\log\mbh$) at fixed $\log\mbh$ ($\log\leddrat$) are shown as solid (dash-dotted) curves. The $25-75\%$ quantiles of $\log\leddrat$ ($\log\mbh$) at fixed $\log\mbh$ ($\log\leddrat$) are shown as dotted (dashed) curves. The error bars represent the $\log\typleddrat$ and $\log\typmbh$ that roughly span the region enclosed by the $25-75\%$ quantiles of $\log\leddrat$ and $\log\mbh$.}\label{fig::flambM}
\end{figure*}

From Equations~\ref{eqn::flambM}, \ref{eqn::phiAGN}, and \ref{eqn::nagn}, we denote $F(\leddrat,\mstar|z)$ as the conditional probability density per unit $\log\leddrat$ per unit $\log\mstar$ that a galaxy at $z$ has \mstar\ and hosts a SMBH accreting at \leddrat:
\begin{equation}\label{eqn::bigF}
\begin{split}
F(\leddrat,\mstar|z)&=\frac{f(\leddrat,\mstar|z)}{\iint f(\leddrat,\mstar|z)\,\D\log\leddrat\D\log\mstar}\\
&=\frac{f(\leddrat,\mstar|z)}{\nagn(z)}.
\end{split}
\end{equation}
The quantity $(1-\epsilon)/(\epsilon c^2) \times F(\leddrat,\mstar|z) \times \mstar \leddrat\kappa(\mstar)$ provides a measure of the contribution to \rhobhar\ per galaxy at different $\log\leddrat$ and $\log\mbh$ in the $\log\leddrat-\log\mbh$ plane after we convert \mstar\ to \mbh\ based upon the adopted $\mbh-\mstar$ relation. Note that the integration in Equation~\ref{eqn::bigF} is for galaxies with $\log\mstar=9.5-12$. Figure~\ref{fig::flambM} shows the results assuming the median \pleddrat, with the integration range of $\log\leddrat$ restricted between --3 and 2. We define $\log\typleddrat$ and $\log\typmbh$ such that they are represented by the intersection point of the running median of $\log\leddrat$ and $\log\mbh$ in the $\log\leddrat-\log\mbh$ plane. The solid error bars represent the regions that are enclosed by the $25-75\%$ quantiles of \leddrat\ and \mbh. Those regions are generally consistent with the maximum values of the heatmaps at all redshifts, indicating our identification of \typleddrat\ and \typmbh\ can truly represent the AGNs that contribute most of \rhobhar.


\begin{figure}[t]
\centering
\includegraphics[width=\linewidth]{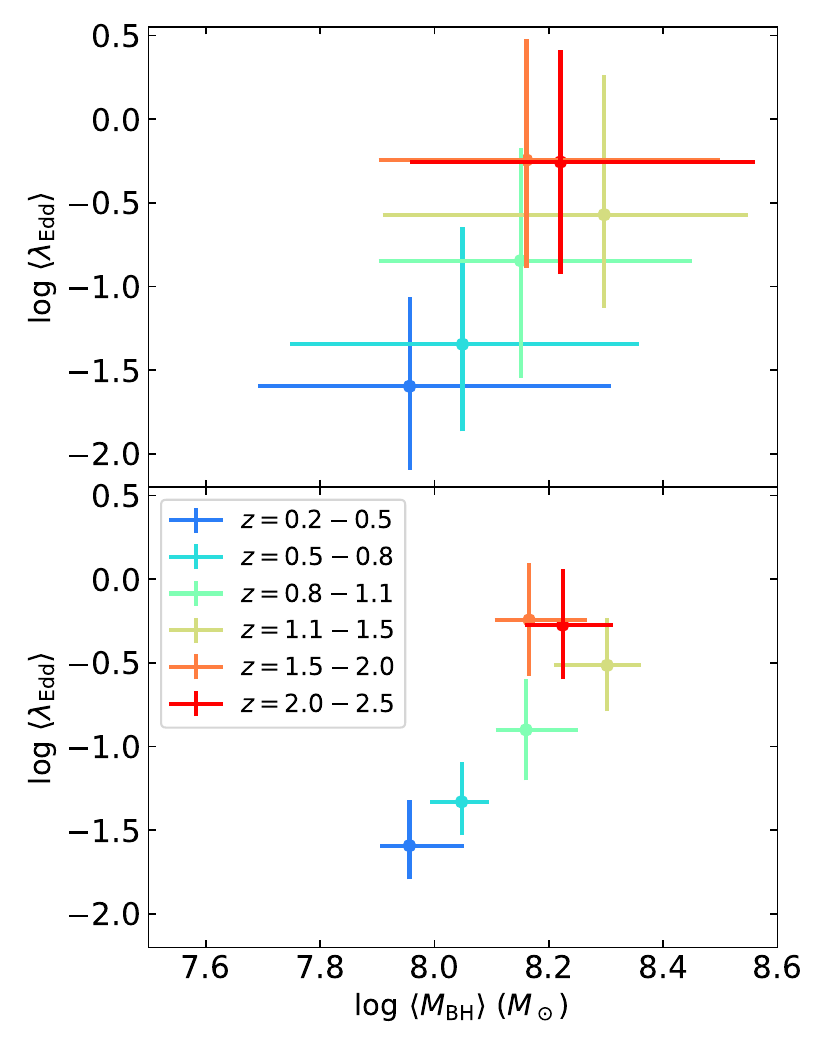}
\caption{The evolution of  $\typleddrat$ and $\typmbh$ at different redshifts, with colors defined in the legend. Top panel: The error bars show the region bounded by the $25-75\%$ quantiles of $\log\leddrat$ and $\log\typmbh$ in Figure~\ref{fig::flambM} assuming the median \pleddrat\ and $\mbh=0.002\mstar$. Bottom panel: The error bars represent the 90\% confidence intervals derived with our Monte Carlo method. From $z=1.5-2.0$ to $z=0.2-0.5$, \typleddrat\ and \typmbh\ both decrease; \typleddrat\ decreases by $1.35\,\rm dex$, but \typmbh\ decreases only by $0.21\,\rm dex$.}\label{fig::typlogleddrat}
\end{figure}

 We further present the evolution of \typleddrat\ and \typmbh\ at different redshifts in Figure~\ref{fig::typlogleddrat}. The error bars in the top panel are the same as those in Figure~\ref{fig::flambM}, while those in the bottom panel represent the 90\% confidence intervals derived with our Monte Carlo method. From Figure~\ref{fig::typlogleddrat}, we can directly quantify the change in \typleddrat\ and \typmbh. At $z=1.5-2.5$, $\log\typleddrat$ is between about $-0.7$ and $0.4$, indicating most of the accretion power is from near-Eddington or super-Eddington AGNs. As redshift decreases, \typleddrat\ decreases significantly with $\Delta\log\typleddrat=-1.35_{-0.39}^{+0.46}\,\mathrm{dex}$ while \typmbh\ only slightly decreases by $0.21_{-0.11}^{+0.11}\,\mathrm{dex}$.\footnote{$\Delta\log\typleddrat=-1.35_{-0.25}^{+0.27}\,\dex$ and $\Delta\log\typmbh=-0.21_{-0.06}^{+0.06}\,\dex$ when considering the 68\% statistical uncertainty.} By $z=0.2$, $\log\typleddrat$ is less than $-1.0$. These trends are consistent with the \rhobhar\ evolution shown in Figure~\ref{fig::rhobhar}, where high-\leddrat\ AGNs dominate most \rhobhar\ at $z\gtrsim1.5$, while low-\leddrat\ AGNs gradually dominate toward lower redshift.



So far, we have quantified \rhobhar, \typleddrat, and \typmbh\ at different redshifts. We now use Equation~\ref{eqn::rho_decompose} to calculate \nagneff, and use Equation~\ref{eqn::nagn} to calculate \nagn\ in different \leddrat\ bins. The results are shown in Figure~\ref{fig::nagn}. We provide simple functional fits to our \nagn\ measurements in different \leddrat\ bins in Appendix~\ref{appedix:functional_fits}. For comparison, in Figure~\ref{fig::nagn} we show the total \nagn\ for AGNs with $\log\lx>42$ from \citet{Buchner+2015}, \citet{Miyaji+2015}, and \citet{Peca+2023}. Our results generally agree with those in the literature. \nagn\ declines at all \leddrat\ from $z\approx2$ to $z\approx0.2$, and AGNs with low-\leddrat\ ($\leddrat<0.1$) always dominate the total \nagn\ at $z<4$. The decline is most significant in the $\leddrat=1-100$ bin, reaching $\approx1.6\,\dex$ from $z=1.5-2.0$ to $z=0.2-0.5$, and the decline in lower-\leddrat\ bins is much smaller, which is consistent with the evolution of \phiMAGN\ shown in Figure~\ref{fig::phiMAGN}. The decline in total \nagn\ is mostly due to the general decline in \leddrat, where AGNs accreting slightly higher than our sample limit ($\leddrat>0.0013$) are shifted below the limit. We have also verified that \nagn\ remain similar after excluding eFEDS from our analyses in Appendix~\ref{appedix:noefeds}. However, \nagneff\ shows dramatically different behavior. From $z=4$ to $z\approx0.5$, \nagneff\ is consistently increasing, and it only shows flattening or a slight downward trend in the last redshift bin of $z=0.2-0.5$. The increase of \nagneff\ from $z=1.5-2.0$ to $z=0.2-0.5$ is $0.29_{-0.46}^{+0.38}\,\mathrm{dex}$.\footnote{$\Delta\log\nagneff=0.29_{-0.27}^{+0.25}\,\dex$ when considering the 68\% uncertainty.} Although the increase is not significant considering its 90\% statistical uncertainty, this result can be understood by the drastic decrease in \typleddrat\ shown in Figure~\ref{fig::typlogleddrat}. As \typleddrat\ decreases, the main contributor to \rhobhar\ is gradually shifted to low-\leddrat\ AGNs that always dominate the total \nagn, but since the total \nagn\ decreases, \nagneff\ is generally constant from $z=1.5-2.0$ to $z=0.2-0.5$. Figure~\ref{fig::nagn} also demonstrates that \nagneff\ is different from the total \nagn. Take the $z=1.5-2$ bin as an example: $\log\typleddrat$ is about between --0.7 and 0.2 in Figure~\ref{fig::typlogleddrat}, which indicates \rhobhar\ is hardly contributed to by AGNs at the other \leddrat\ values. Thus \nagneff\ is similar to \nagn\ in the $\leddrat=0.1-1$ bin (as shown in Figure~\ref{fig::nagn}) and hardly counts AGNs in the other \leddrat\ bins. We also calculate the expected \nagn\ for AGNs within the $25-75\%$ quantile of $\leddrat$ (i.e., for AGNs with \typleddrat). The estimates for \nagneff\ and \nagn\ for AGNs with \typleddrat\ are generally consistent within the 90\% statistical uncertainty, indicating that our estimated \typleddrat\ can truly represent the typical \leddrat.

\centerwidetable
\begin{deluxetable*}{lcccccccccc}
\tabletypesize{\footnotesize}
\tablewidth{0pt}
\tablecaption{$\log\rhobhar$, $\log\typleddrat$, $\log\typmbh$, and $\log\nagneff$ at different redshifts bins and their declines.
\label{tab::decline}}
\tablehead{
\colhead{$z$} & \colhead{$0.2-0.5$} & \colhead{$0.5-0.8$} & \colhead{$0.8-1.1$} & \colhead{$1.1-1.5$} & \colhead{$1.5-2.0$} & \colhead{$2.0-2.5$} & \colhead{$2.5-3.0$} & \colhead{$3.0-3.5$} & \colhead{$3.5-4.0$} & \colhead{$\Delta$ (dex)}
}
\startdata
\hline
$\log\rhobhar$ & $-5.83_{-0.06}^{+0.07}$ & $-5.32_{-0.05}^{+0.05}$ & $-4.83_{-0.04}^{+0.05}$ & $-4.58_{-0.04}^{+0.05}$ & $-4.54_{-0.05}^{+0.05}$ & $-4.61_{-0.07}^{+0.06}$ & $-4.86_{-0.08}^{+0.08}$ & $-4.75_{-0.09}^{+0.09}$ & $-5.33_{-0.17}^{+0.17}$ & $-1.28_{-0.08}^{+0.08}$ \\
$\log\typleddrat$ & $-1.59_{-0.20}^{+0.27}$ & $-1.33_{-0.20}^{+0.24}$ & $-0.90_{-0.30}^{+0.30}$ & $-0.51_{-0.27}^{+0.28}$ & $-0.24_{-0.33}^{+0.34}$ & $-0.27_{-0.32}^{+0.34}$ & $-0.16_{-0.41}^{+0.40}$ & $0.20_{-0.35}^{+0.34}$ & $0.35_{-0.67}^{+0.63}$ & $-1.35_{-0.39}^{+0.46}$ \\
$\log\typmbh$ & $\phantom{X}7.96_{-0.05}^{+0.10}$ & $\phantom{X}8.05_{-0.05}^{+0.05}$ & $\phantom{X}8.16_{-0.05}^{+0.09}$ & $\phantom{X}8.30_{-0.09}^{+0.06}$ & $\phantom{X}8.17_{-0.06}^{+0.10}$ & $\phantom{X}8.22_{-0.07}^{+0.09}$ & $\phantom{X}8.15_{-0.14}^{+0.16}$ & $\phantom{X}8.26_{-0.10}^{+0.15}$ & $\phantom{X}7.89_{-0.15}^{+0.16}$ & $-0.21_{-0.11}^{+0.11}$ \\
$\log\nagneff$ & $-4.50_{-0.29}^{+0.21}$ & $-4.33_{-0.20}^{+0.19}$ & $-4.40_{-0.27}^{+0.23}$ & $-4.66_{-0.29}^{+0.26}$ & $-4.78_{-0.34}^{+0.32}$ & $-4.87_{-0.38}^{+0.35}$ & $-5.13_{-0.51}^{+0.46}$ & $-5.54_{-0.34}^{+0.42}$ & $-5.88_{-0.61}^{+0.71}$ & $\phantom{X}0.29_{-0.46}^{+0.38}$ \\
\hline
$\log\rhobhar$ & $-5.81_{-0.06}^{+0.06}$ & $-5.32_{-0.05}^{+0.05}$ & $-4.84_{-0.05}^{+0.04}$ & $-4.59_{-0.05}^{+0.05}$ & $-4.56_{-0.05}^{+0.05}$ & $-4.62_{-0.06}^{+0.06}$ & $-4.88_{-0.07}^{+0.08}$ & $-4.76_{-0.08}^{+0.09}$ & $-5.36_{-0.16}^{+0.16}$ & $-1.26_{-0.08}^{+0.08}$ \\
$\log\typleddrat$ & $-1.45_{-0.26}^{+0.29}$ & $-1.20_{-0.23}^{+0.30}$ & $-0.83_{-0.31}^{+0.37}$ & $-0.49_{-0.26}^{+0.35}$ & $-0.16_{-0.51}^{+0.45}$ & $-0.23_{-0.34}^{+0.36}$ & $-0.04_{-0.43}^{+0.57}$ & $0.27_{-0.36}^{+0.40}$ & $0.59_{-0.66}^{+0.70}$ & $-1.28_{-0.53}^{+0.58}$ \\
$\log\typmbh$ & $\phantom{X}7.80_{-0.19}^{+0.22}$ & $\phantom{X}7.90_{-0.13}^{+0.14}$ & $\phantom{X}8.06_{-0.11}^{+0.21}$ & $\phantom{X}8.28_{-0.17}^{+0.10}$ & $\phantom{X}8.05_{-0.10}^{+0.23}$ & $\phantom{X}8.20_{-0.10}^{+0.14}$ & $\phantom{X}8.03_{-0.25}^{+0.18}$ & $\phantom{X}8.18_{-0.15}^{+0.18}$ & $\phantom{X}7.61_{-0.24}^{+0.25}$ & $-0.25_{-0.31}^{+0.26}$ \\
$\log\nagneff$ & $-4.49_{-0.16}^{+0.16}$ & $-4.33_{-0.16}^{+0.13}$ & $-4.38_{-0.27}^{+0.16}$ & $-4.67_{-0.31}^{+0.18}$ & $-4.76_{-0.40}^{+0.34}$ & $-4.88_{-0.37}^{+0.33}$ & $-5.17_{-0.49}^{+0.40}$ & $-5.53_{-0.35}^{+0.37}$ & $-5.89_{-0.63}^{+0.73}$ & $\phantom{X}0.27_{-0.39}^{+0.44}$ \\
\hline
$N_\mathrm{X}$ & 873 & 1198 & 1367 & 1570 & 1281 & 563 & 246 & 136 & 34 & Total: 7268 \\
$N_\mathrm{gal}$ & 325578 & 360191 & 253990 & 186904 & 105648 & 40967 & 22843 & 12252 & 4919 & Total: 1313292 \\
\enddata
\tablecomments{The first part of the table shows the results assuming $\mbh=0.002\,\mstar$, while the second part adopts the relation $\log\mbh=7.43+1.61\log(\mstar/M_0)$ with $M_0=3\times10^{10}\msun$ \citep{Greene+2020}. The uncertainties represent the 90\% confidence intervals derived from 1000 Monte Carlo trials. ``$\Delta$" indicates the difference between the $z=0.2-0.5$ and $z=1.5-2.0$ bins, with negative values indicating a decline toward lower redshifts. The last two rows show the number of \xray\ AGNs ($N_\mathrm{X}$) and the number of normal galaxies ($N_\mathrm{gal}$) above the \mstar-completeness limits.}
\end{deluxetable*}

\begin{figure}[ht]
\centering
\includegraphics[width=\linewidth]{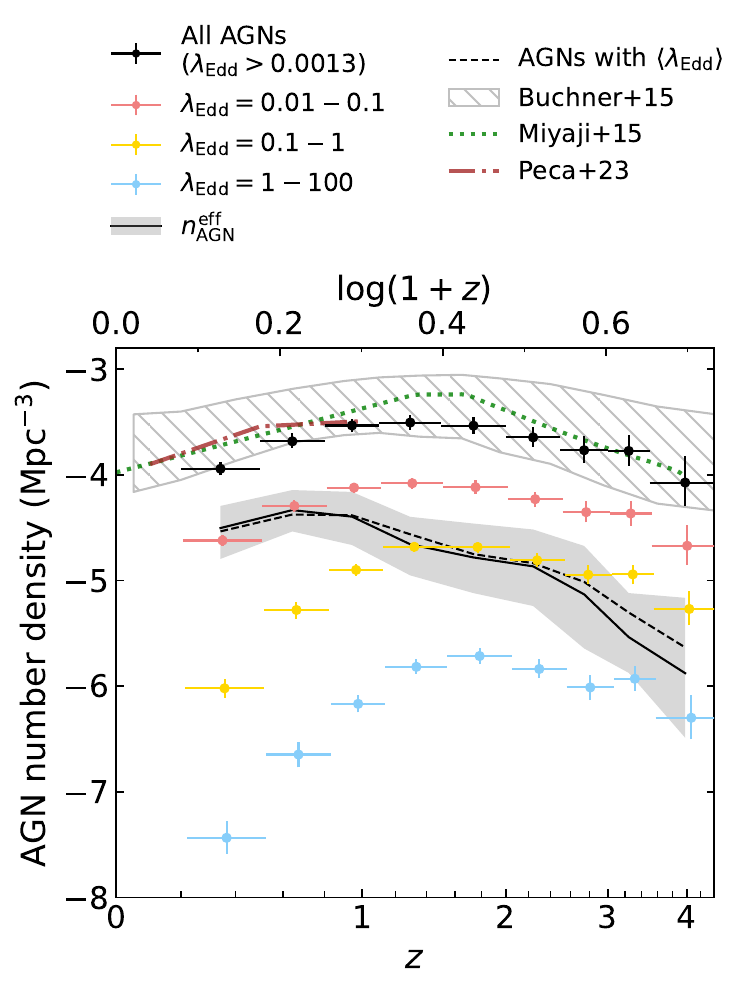}
\caption{\nagn\ and \nagneff\ as a function of redshift. The black data points represent the total \nagn\ sampled by our data ($\leddrat>0.0013$). The red, yellow, and blue data points represent \nagn\ for AGNs accreting at $\leddrat=0.01-0.1$, $\leddrat=0.1-1$, and $\leddrat=1-100$, respectively. The black solid line represents \nagneff, and the dashed line represents \nagn\ for AGNs within the $25-75\%$ quantile range of \typleddrat. The error bars of the data points and the grey shaded region for \nagneff\ represent the 90\% confidence intervals derived with our Monte Carlo method. For comparison, we show the total \nagn\ for AGNs with $\log\lx>42$ from \citet{Buchner+2015} (grey hatched region representing 10--90\% confidence intervals), \citet{Miyaji+2015} (green dotted line), and \citet{Peca+2023} (brown dotted-dashed line).}\label{fig::nagn}
\end{figure}

We summarize our results in Table~\ref{tab::decline}. We also test the $\mbh-\mstar$ relation in \citet{Greene+2020}, which is nonlinear with $\mbh\propto\mstar^{1.61}$. The results are generally similar to those assuming $\mbh=0.002\,\mstar$.

From our analyses above, we have shown that the decline in \typleddrat\ is the primary driver of the broad decline in \rhobhar. The contribution from \typmbh\ ($\Delta\log\typmbh=-0.21_{-0.11}^{+0.11}\,\mathrm{dex}$) is about 13 times smaller than that from \typleddrat\ ($\Delta\log\typleddrat=-1.35_{-0.39}^{+0.46}\,\mathrm{dex}$). On the other hand, although \nagn\ for AGNs in all \leddrat\ bins drops significantly, \nagneff\ increases because it is driven by the decline in \typleddrat. Our results are also generally consistent with and extend those in \citet{Nandra+2025}, where they compare the low-redshift (median $z=0.34$) eFEDS hard \xray\ selected AGNs with the higher-redshift sample from the Chandra COSMOS legacy survey of \citet{Suh+2020} (median $z=1.58$). These two samples have similar \mbh\ distributions and luminosity limits at their median redshifts, but the eFEDS low-redshift sample has significantly lower median \leddrat, indicating the decline in global accretion rate is primarily due to a reduction in the typical \leddrat, rather than a shift of accretion to lower-mass SMBHs.


Our analyses assume no significant redshift evolution in the $\mbh-\mstar$ relation at $z<2$ as supported by the recent studies of, e.g., \citet{Suh+2020} and \citet{Li+2023}. However, some earlier studies report an increasing $\mbh/\mstar$ ratio toward higher redshift \citep[$\approx0.3\,\dex$ from $z\approx0$ to $z\approx2$; e.g.,][]{Merloni+2010, Trakhtenbrot+2010}. As a basic estimate of the impact of such evolution, \typleddrat\ decreases by $\approx0.3\,\dex$ and \typmbh\ increases by $\approx0.3\,\dex$ at $z=2.0-2.5$, while the results at $z=0.2-0.5$ remain unchanged in Figure~\ref{fig::typlogleddrat}. Thus, $\Delta\log\typleddrat\approx-1.0\,\dex$ and $\Delta\log\typmbh\approx-0.5\,\dex$ from $z\approx2$ to $z\approx0.2$. This estimate does not alter our overall conclusion that the decline in \typleddrat\ is the primary driver of the broad decline in \rhobhar. It should also be noted that an evolving observed $\mbh/\mstar$ ratio does not necessarily imply evolution of the underlying $\mbh-\mstar$ relation if selection bias is not properly taken into account. For example, in the $\mbh\propto\mstar^{1.61}$ relation of \citet{Greene+2020}, selection bias may suppress the detection of lower-mass galaxies at higher redshift, leading to an apparently higher $\mbh/\mstar$ ratio ($\mbh/\mstar\propto\mstar^{0.61}$) even if the $\mbh-\mstar$ relation itself does not evolve. In this case, our results would not differ significantly, as shown in Table~\ref{tab::decline}.

One systematic bias arises from potentially missed CT accretion. 
However, as discussed in Section~\ref{subsec::data}, this appears to affect \rhobhar\ by $\lesssim0.2\,\dex$ over the $z\lesssim2$ range applicable to this paper. 
Another source of systematic uncertainty involves \kbol. We tested the \kbol\ from \citet{Yang+2018} \citep[a modification of][]{Lusso+2012} and found no material differences in the results. Additionally, while \typleddrat\ and \typmbh\ lack universally precise definitions, their determination is supported by the consistency between \nagneff\ and \nagn\ for AGNs with \typleddrat\ in Figure~\ref{fig::nagn}, indicating that our methodology is robust.

Although our results robustly reveal the primary cause for the decline in SMBH growth at $z\lesssim2$ thanks to the well-constrained \pleddrat, we do not attempt to explain the rise in SMBH growth from $z\approx4$ to $z\approx2$ with the same methodology. This is mainly because of the higher obscured AGN fraction at $z>2$ and the emerging evidence that there may be more hidden accretion power than previously expected. It is found that the obscured AGN fraction in \xray s appears to increase with redshift \citep[e.g.,][]{Buchner+2015,Liu+2017,Vito+2018,Lyu+2024}. Also, new JWST results indicate that some AGNs may be missed by \xray\ surveys, even after accounting for observational biases. For example, many apparent \xray\ weak AGNs were discovered by JWST at $z\gtrsim2$ \citep[e.g.,][]{Kocevski+2025,Maiolino+2025}. These AGNs may be heavily obscured in \xray s or intrinsically \xray\ weak. \citet{Yang+2021} showed that the \xray-inferred \rhobhar\ may be underestimated by a factor of a few at $z>3$. Such a discrepancy mainly occurs at high redshift and does not have a significant impact on our analyses of the decline at $z\lesssim2$. It is worthwhile to explore the cause of this discrepancy and quantify the impact of the higher $\mbh/\mstar$ ratio observed among JWST-selected \xray-weak AGNs \citep[e.g.,][]{Kocevski+2025} on \rhobhar\ at $z>3$, but such efforts are beyond the scope of this paper.



\subsection{Does \mstar\ Mainly Modulate the Typical Outburst Luminosity or Duty Cycle to Reduce SMBH Growth?}\label{subsec::mstar}

In this subsection, we investigate the decline in \bhar\ as \mstar\ decreases at fixed redshift, as per key question~2. 
With a similar approach to that in Section~\ref{subsec::decline_in_z}, \bhar\ can be approximately factored into two components and a constant factor at fixed (\mstar, $z$):
\begin{equation}
    \bhar=\lbolagn\times\fagn\times\frac{(1-\epsilon)\,\kappa(\mstar)}{\epsilon c^2},
\end{equation}
where \lbolagn\ is the sample-averaged outburst luminosity \textit{per AGN} and \fagn\ is the AGN duty cycle (or equivalently the AGN fraction). Note that our \bhar\ is based upon the sample-averaged \lbol\ \textit{per galaxy}, which is different from \lbolagn. \mstar\ can modulate \lbolagn\ or \fagn\ to change \bhar.

We can calculate the \fagn\ for galaxies if we regard SMBHs in those galaxies accreting above a given \leddrat\ (\leddratmin) as AGNs:
\begin{equation}\label{eqn::fagn}
\fagn=\int_{\log\leddratmin}^{+\infty}\pleddrat\D\log\leddrat.
\end{equation}
On the other hand, the \lbolagn\ can be calculated as
\begin{equation}
\begin{split}
\lbolagn=\int_{\log\leddratmin}^{+\infty}\mstar&\leddrat\kappa(\mstar)\\ &\times\pleddrat/\fagn\,\D\log\leddrat.
\end{split}
\end{equation}                               
The term $\pleddrat/\fagn$ represents the conditional probability that \textit{an AGN} with \mstar\ and $z$ accretes at \leddrat. We set $\leddratmin=0.01$ following \citet{Aird+2018}.

\begin{figure*}[htb]
\centering
\includegraphics[width=\linewidth]{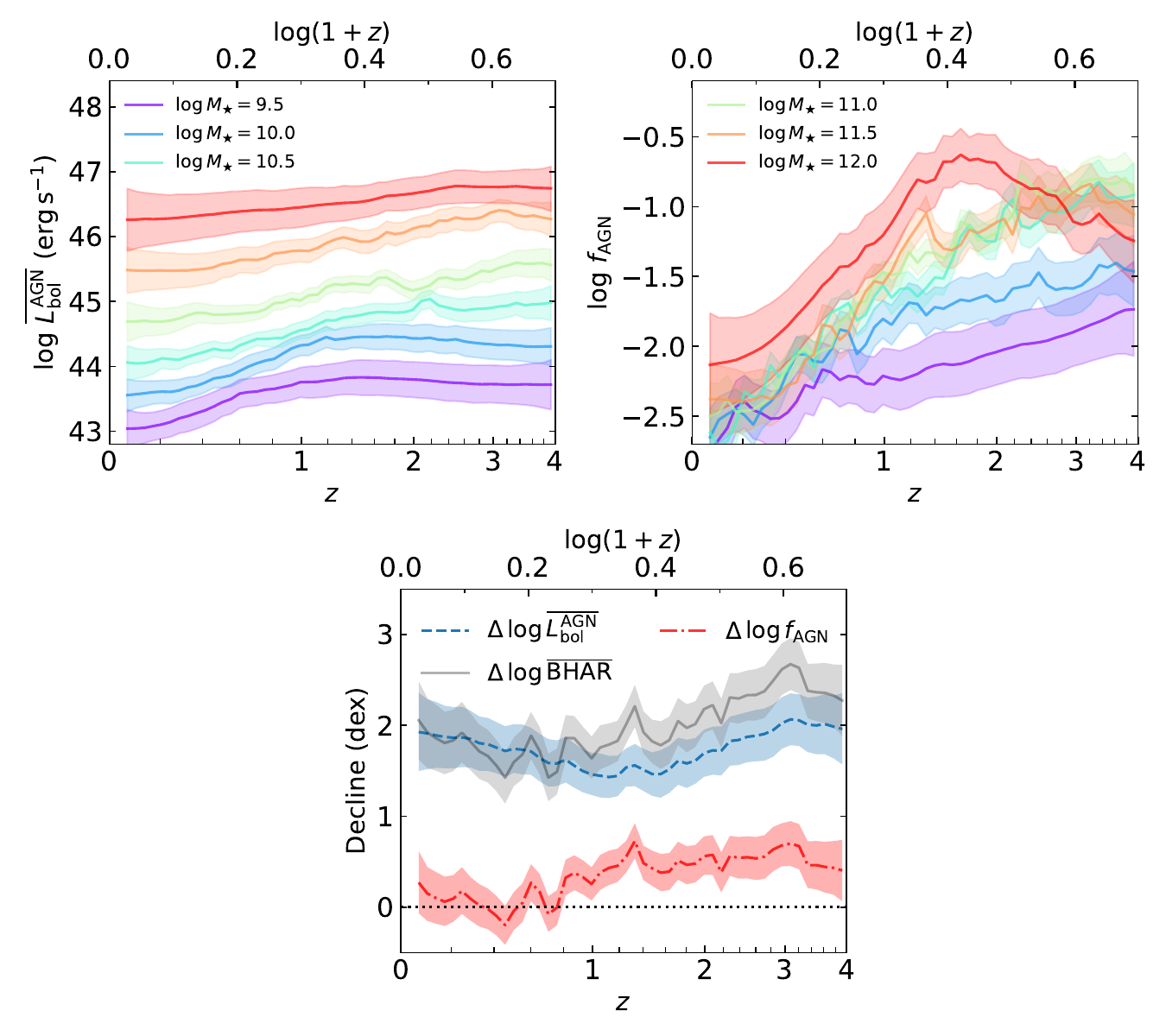}
\caption{Top panels: \lbolagn\ (top-left) and \fagn\ (top-right) as functions of redshift at different \mstar, with colors defined in the legend. Bottom panel: The decline in \bhar\ (grey solid), \lbolagn\ (blue dashed), and \fagn\ (red dash-dotted) from $\log\mstar=11.5$ to $\log\mstar=10$ as functions of redshift. The colored shaded stripes represent the $1\sigma$ uncertainty from \pleddrat. }\label{fig::Lbol_and_fagn}
\end{figure*}

The top panels of Figure~\ref{fig::Lbol_and_fagn} show \lbolagn\ and \fagn\ as functions of redshift at different \mstar. \mstar\ strongly modulates \lbolagn\ at all redshifts. This behavior generally reflects the $\mbh-\mstar$ relation, where more massive galaxies host more massive SMBHs. Without a strict dependence of \leddrat\ on \mstar\ at fixed redshift, more massive galaxies tend to host more luminous AGNs.
\lbolagn\ also slightly decreases toward lower redshift at fixed \mstar. With our assumption of a linear scaling between \mbh\ and \mstar, this result indicates a slight decrease in \leddrat. The small decrease in \leddrat\ here does not contradict the large evolution of \typleddrat\ in Section~\ref{subsubsec::impact_of_three}, because \typleddrat\ characterizes the typical \leddrat\ for the AGNs contributing most of \rhobhar, while \lbolagn\ is simply the average \lbol\ of AGNs. \nagn\ is dominated by low-\leddrat\ AGNs at all redshifts as shown in Figure~\ref{fig::nagn}, causing a less-significant evolution of the average \leddrat. On the other hand, in Figure~\ref{fig::Lbol_and_fagn} top-right panel, \fagn\ is almost independent of \mstar\ at low redshift and is clearly dependent on \mstar\ only at $z\gtrsim1$. At fixed \mstar, \fagn\ show stronger redshift evolution in more massive galaxies. Such behavior for \fagn\ has been found in previous works \citep[e.g.,][]{Aird+2018,Aird+2019,Birchall+2022}. At $z\gtrsim1$, larger amounts of gas are available \citep[e.g.,][]{Decarli+2019}, which tends to trigger luminous AGNs in massive galaxies with massive SMBHs and deeper potential wells \citep[e.g.,][]{Rosas-Guevara+2015}. As redshift decreases, the lack of available gas starves luminous AGNs, causing a more significant decline in \fagn\ in massive galaxies.



The bottom panel of Figure~\ref{fig::Lbol_and_fagn} shows the declines in \bhar, \lbolagn, and \fagn\ from $\log\mstar=11.5$ to $\log\mstar=10$ as functions of redshift. These two $\log\mstar$ values are chosen for illustration because our measurements at these masses are well-constrained. At $z\lesssim1$, the decline in \bhar\ is driven almost entirely by \lbolagn. At $z\gtrsim1$, the decline in \fagn\ is only about $0.3-0.7\,\dex$, which is more than 10 times smaller than the decline in \lbolagn\ ($1.4-2.0\,\dex$).

Overall, our results indicate that \mstar\ predominantly regulates \lbolagn\ instead of \fagn\ to drive variations in \bhar. These results are also generally consistent with those from \citet{Aird+2018} in their Figures~6 and 8, where higher-\mstar\ galaxies exhibit a higher \fagn\ than lower-\mstar\ galaxies at $z\gtrsim1$, and the decrease in \lbolagn\ for lower-\mstar\ galaxies is slightly faster than that for higher-\mstar\ galaxies. Our measurements provide superior constraints for a wide \mstar\ range than those from \citet{Aird+2018} thanks to the larger survey volume of XMM-SERVS and eFEDS. We also test the nonlinear $\mbh-\mstar$ relation of \citet{Greene+2020} in our analyses and find an even higher fractional contribution of \lbolagn\ to the variation in \bhar. This arises because the nonlinear relation ($\mbh\propto\mstar^{1.61}$) imposes a stronger \mstar-dependence than the linear case, yielding larger differences in \mbh\ between galaxies of different \mstar, and thus enhancing the contribution of \lbolagn\ in driving the variation in \bhar.

\section{Summary and Future Work} \label{sec::summary}

In this work, we leverage the best-measured sample-averaged SMBH accretion rates in \citet{Zou+2024a} to understand and quantify the decline in SMBH growth at $z\lesssim2$. Our main results are summarized as follows:
\begin{enumerate}
\item We confirm that \rhobhar\ peaks at $z\approx2$ and declines dramatically since then. We find that low-accretion activity gradually dominates the contribution to \rhobhar\ as redshift decreases. At $z\gtrsim1$, \rhobhar\ is mostly contributed by rapidly accreting SMBHs, while at $z\lesssim0.5$, $\leddrat<0.1$ activity dominates \rhobhar. See Section~\ref{subsubsec::rhobhar}.

\item From $z\approx2$ to $z\approx0.2$, the decline in \rhobhar\ is mainly driven by the decline in \leddrat\ ($\Delta\log\typleddrat=-1.35_{-0.39}^{+0.46}\,\dex$) rather than \mbh\ ($\Delta\log\typmbh=-0.21_{-0.11}^{+0.11}\,\dex$). On the other hand, \nagneff\ is generally constant (increasing by $0.29_{-0.46}^{+0.38}\,\dex$) due to the significantly lower \typleddrat\ at lower redshift, the decreasing total \nagn, and the fact that the low-\leddrat\ AGN number density dominates the total \nagn\ at all redshifts. See Section~\ref{subsubsec::impact_of_three}.

\item The observed dependence of \bhar\ on \mstar\ at fixed redshift arises primarily because \mstar\ predominantly regulates \lbolagn\ instead of \fagn\ to drive variations in \bhar. \fagn\ shows stronger redshift evolution in more massive galaxies, which may be due to the availability of cold gas preferentially affecting luminous AGNs in massive galaxies.
See Section~\ref{subsec::mstar}.
\end{enumerate}

Overall, our results clarify the primary cause of the decline in SMBH growth at $z<2$ and provide new insight into the AGN downsizing phenomenon. Previous studies suggest that AGN downsizing does not reflect antihierarchical behavior, but instead results from a combination of factors, e.g., \leddrat\ and \mbh, where \leddrat\ may play a more important role \citep[e.g.,][]{Babic+2007, Fanidakis+2012, Aird+2015, Suh+2015}. Our findings provide the most compelling evidence to date that \leddrat\ plays the primary role in driving AGN downsizing.

There are several ways to extend this work with better datasets in the future. For example, there are existing wide-field \xray\ surveys by Chandra and/or XMM-Newton with generally sufficient quality multiwavelength coverage, such as the Chandra Deep Wide-Field Survey \citep[\mbox{CDWFS}; e.g.,][]{Masini+2020}, XMM-XXL \citep[e.g.,][]{Pierre+2016}, Stripe~82X \citep[e.g.,][]{LaMassa+2016}, and Stripe~82-XL \citep[e.g.,][]{Peca+2024}, totaling $\sim100\,\deg^2$. Future work could include these surveys in the ``wedding-cake" design to help sample AGNs at low redshift and/or high luminosity. There are also other very wide-field \xray\ catalogs above $2\,\kev$—such as the Chandra Source Catalog \citep[e.g.,][]{Evans+2024}, 4XMM \citep[e.g.,][]{Webb+2020}, and the Swift/BAT catalog \citep[e.g.,][]{Lien+2025}—providing a much larger survey volume with reduced obscuration effects, particularly valuable for sampling the most luminous AGNs. Soon, these catalogs will benefit from excellent multiwavelength coverage provided by new photometric and spectroscopic surveys such as LSST \citep[e.g.,][]{Ivezic+2019}, Euclid \citep[e.g.,][]{EuclidCollaboration+2024}, Roman \citep[e.g.,][]{Akeson+2019}, Wide Field Survey Telescope \citep[WFST; e.g.,][]{Wang+2023}, Chinese Space Station Survey Telescope \citep[CSST; e.g.,][]{CSSTCollaboration+2025}, SPHEREx \citep[e.g.,][]{Crill+2020}, 4\,m Multi-Object Spectroscopic Telescope \citep[4MOST; e.g.,][]{deJong+2019}, Subaru Prime Focus Spectrograph \citep[PFS; e.g.,][]{Takada+2014}, and Dark Energy Spectroscopic Instrument \citep[DESI; e.g.,][]{DESICollaboration+2025}. These complementary datasets will allow the construction of massive samples of well-characterized \xray\ AGNs with robust redshifts and host-galaxy properties. Also, future deep X-ray surveys by the NewAthena \citep[e.g.,][]{Cruise+2025}, AXIS \citep[e.g.,][]{Reynolds+2023}, and Lynx \citep[e.g.,][]{Gaskin+2019} missions can probe AGN populations with higher obscuration, which can constrain the missed population caused by obscuration.

\begin{acknowledgments}

We thank the anonymous referee for constructive feedback. ZY and WNB acknowledge support from NSF grants AST-2106990 and AST-2407089 and Chandra X-ray Center grant AR4-25008X. FV acknowledges support from ``INAF Ricerca Fondamentale 2023 - Large GO" grant.

\end{acknowledgments}

\appendix

\begin{figure*}[ht]
\centering
\includegraphics[width=\linewidth]{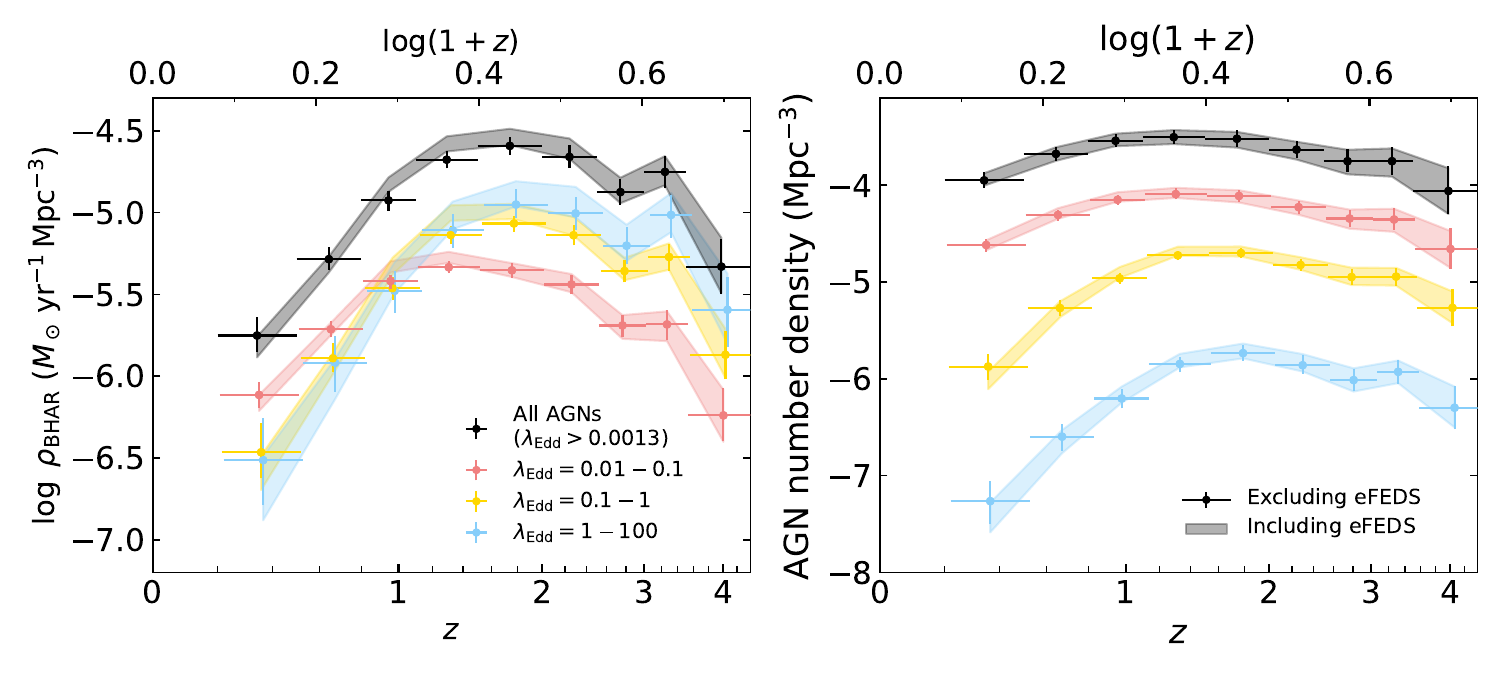}
\caption{Comparison of \rhobhar\ (left panel) and \nagn\ (right panel) for results excluding eFEDS (data points with error bars) and including eFEDS (shaded regions). Error bars and shaded regions indicate 90\% statistical uncertainties, with colors matching those in the legend of the left panel.}\label{fig::noefeds}
\end{figure*}

\section{Results without eFEDS}\label{appedix:noefeds}

eFEDS is mainly observed in soft \xray s below $2.3\,\kev$, which makes it more easily affected by obscuration (though, as discussed in Section~\ref{subsec::data}, corrections have been made for obscured accretion power). In this Appendix, we test if eFEDS introduces significant bias by excluding eFEDS from our analyses, and our main results in Section~\ref{subsec::decline_in_z} are shown in Figure~\ref{fig::noefeds}. After excluding eFEDS, the results for \rhobhar\ and \nagn\ in different \leddrat\ bins generally exhibit larger uncertainties at $z\lesssim1$, but their median values remain consistent with those including eFEDS within the $90\%$ confidence intervals. These results demonstrate that eFEDS does not lead to significant bias, nor does excluding eFEDS alter the overall conclusion that the broad decline in \rhobhar\ is mainly driven by the decline in \typleddrat. In addition to the other fields covering $\approx13\,\deg^2$ that are observed from $\approx2-10\,\kev$, the $60\,\deg^2$ eFEDS field provides useful constraints at $z\lesssim1$ by increasing the number of \xray\ AGNs at $z\lesssim1$ by $\approx60\%$, thereby allowing more accurate measurements of the SMBH growth decline at $z\lesssim2$.


\section{Functional Fits to the redshift evolution of $\rhobhar$ and $\nagn$}\label{appedix:functional_fits}

In Section~\ref{subsec::decline_in_z}, we show that \rhobhar\ and \nagn\ in different redshift and \leddrat\ bins can be calculated using Equations~\ref{eqn::rhobhar} and \ref{eqn::nagn}, respectively. In this Appendix, we present simple functional fits to their redshift evolution at $z<4$ using a smoothed double power-law:
\begin{equation}
    \rhobhar(z)\ \mathrm{or}\ \nagn(z)=A\left[\left(\frac{1+z}{1+z_0}\right)^{\gamma_1} + \left(\frac{1+z}{1+z_0}\right)^{\gamma_2}\right]^{-1},
\end{equation}
where $A$ is the normalization, $z_0$ is approximately the redshift where \rhobhar\ or \nagn\ peaks, and $\gamma_1$ and $\gamma_2$ are the different slopes before and after $z_0$, as seen in our Figures~\ref{fig::rhobhar} and \ref{fig::nagn}. Such a functional form is similar to the broken power-law shape of the redshift evolution term adopted in the luminosity-dependent density evolution (LDDE) model \citep[e.g.,][]{Ueda+2014,Aird+2015,Buchner+2015,Pouliasis+2024}. However, we do not consider any \leddrat\ or \lx\ dependence, and we fit \rhobhar\ and \nagn\ in different \leddrat\ bins separately. We use a least-squares method to fit our median data points assuming the linear $\mbh-\mstar$ relation, and the best-fit parameters are summarized in Table~\ref{tab::simple_fit}.

\centerwidetable
\begin{deluxetable*}{ccccccccc}\label{tab::simple_fit}
\tabletypesize{\footnotesize}
\tablewidth{0pt}
\tablecaption{Smoothed double power-law fits to the redshift evolution of \rhobhar\ and \nagn.
\label{tab::fitting}}
\tablehead{
\colhead{} & \colhead{\rhobhar} & \colhead{} & \colhead{} & \colhead{} & \colhead{\nagn} & \colhead{} & \colhead{} & \colhead{}}
\startdata
& All AGNs & $\leddrat=0.01-0.1$ & $\leddrat=0.1-1$ & $\leddrat=1-100$ &  All AGNs & $\leddrat=0.01-0.1$ & $\leddrat=0.1-1$ & $\leddrat=1-100$\\
$\log A$ & $-4.20 \pm 0.09$ & $-4.93 \pm 0.07$ & $-4.64 \pm 0.09$ & $-4.59 \pm 0.11$ & $-3.20 \pm 0.03$ & $-3.78 \pm 0.04$ & $-4.42 \pm 0.06$ & $-5.49 \pm 0.07$ \\
$z_0$ & \phantom{X}$1.48 \pm 0.21$ & \phantom{X}$1.31 \pm 0.20$ & \phantom{X}$1.39 \pm 0.15$ & \phantom{X}$1.48 \pm 0.18$ & \phantom{X}$1.18 \pm 0.22$ & \phantom{X}$1.12 \pm 0.18$ & \phantom{X}$1.13 \pm 0.09$ & \phantom{X}$1.29 \pm 0.11$ \\
$\gamma_1$ & \phantom{X}$3.30 \pm 0.83$ & \phantom{X}$3.56 \pm 0.68$ & \phantom{X}$3.39 \pm 0.73$ & \phantom{X}$2.85 \pm 0.85$ & \phantom{X}$2.29 \pm 0.40$ & \phantom{X}$2.24 \pm 0.36$ & \phantom{X}$2.09 \pm 0.33$ & \phantom{X}$2.16 \pm 0.45$ \\
$\gamma_2$ & $-6.07 \pm 0.99$ & $-5.15 \pm 1.07$ & $-7.77 \pm 1.02$ & $-7.86 \pm 1.03$ & $-3.36 \pm 0.73$ & $-4.07 \pm 0.75$ & $-7.84 \pm 0.74$ & $-8.23 \pm 0.74$ \\
\hline
\enddata
\tablecomments{The second through fifth columns and the last four columns show the best-fit results for \rhobhar\ and \nagn, respectively. The uncertainties represent $1\sigma$ confidence intervals.}
\end{deluxetable*}

\bibliography{sample7}
\bibliographystyle{aasjournalv7}


\end{CJK*}
\end{document}